\documentstyle[psfig]{cupconf}

\ifoldfss
\else
  \ifnfssone
    \newmathalphabet{\mathit}
      \addtoversion{normal}{\mathit}{cmr}{m}{it}
      \addtoversion{bold}{\mathit}{cmr}{bx}{it}
    \newmathalphabet{\mathcal}
      \addtoversion{normal}{\mathcal}{cmsy}{m}{n}
    \else
    \ifnfsstwo
    \fi
  \fi
\fi

\def\hexnumber#1{\ifcase#1 0\or1\or2\or3\or4\or5\or6\or7\or8\or9\or
 A\or B\or C\or D\or E\or F\fi }

\makeatletter
\ifx\CUP@mtlplain@loaded\undefined
\else
\fi
\makeatother
\makeatletter
\ifx\CUP@mtlplain@loaded\undefined
\else
\fi
\makeatother
\title[Evolution of primordial $H_2$]{Evolution of primordial $H_2$ 
for different cosmological models
\footnote{Talk given at the International Conference {\it $H_2$ in space}, 
IAP-Paris (France)\\
 September 28$^{th}$-October 1$^{st}$, 1999}
}

\author[D. PUY]
{Denis PUY}

\affiliation{Paul Scherrer Institut, Laboratory for Astrophysics, 5232 Villigen (Switzerland)\\
Institute of Theoretical Physics, University of Zurich, 8057 Zurich (Switzerland)}

\begin{document}
\ifnfssone
\else
  \ifnfsstwo
  \else
    \ifoldfss
      \let\mathcal\cal
      \let\mathrm\rm
      \let\mathsf\sf
    \fi
  \fi
\fi

\maketitle

\begin{abstract}
Primordial chemistry began in the recombination epoch when the adiabatic 
expansion caused the temperature of the radiation to fall below 4000 K. The 
chemistry of the early Universe involves the elements hydrogen, its isotope 
deuterium, helium with its isotopic forms and lithium. 
\\
In this talk I will present results on the evolution of the primordial $H_2$ 
abundance for different cosmological models and the influence 
on the thermal decoupling.
\end{abstract}

\firstsection 
\section{Introduction}
At early times the Universe was filled up with an extremely dense and hot gas. 
Due to the expansion it cooled below the binding energies of hydrogen, 
deuterium, helium, lithium, and thus one can expect the formation 
of these nuclei. As soon as neutrons and protons leave the equilibrium, the 
formation of deuterium followed by fast reactions lead finally to the formation
 of tritium and helium. 
Thus deuterium is the first stone of the nucleosynthesis but 
also the {\it passage oblig\'e} for heavier elements such as lithium, 
beryllium and boron. 
The basic conclusions of the big bang nucleosynthesis on the baryon density 
$\Omega_\rho$ are
\begin{equation}
0.01 \, < \, \Omega_{\rho} h^2 \, < \, 0.025
\end{equation}
See Sarkar (1996), Olive (1999) and references in Signore \& Puy (1999).

\section{Post-recombination chemistry}
The study of chemistry in the post recombination epoch has grown considerably 
in recent years. From the pionneer works of Saslaw \& Zipoy (1967), Shchekinov 
\& Ent\'el (1983), Lepp \& Shull (1987), Dalgarno \& Lepp (1987) and Black 
(1988), many authors have developped studies of primordial chemistry in 
different contexts. Latter \& Black (1991), Puy et al. (1993), Stancil et al. 
(1996) for the chemical network and the thermal balance , 
Palla, Galli \& Silk (1995), Puy \& Signore 
(1996, 1997, 1998a, 1998b), Abel et al. (1997) and Galli \& Palla (1998) for 
the study of the initial conditions of the formation of the first objects. 

\subsection{History}

From the recombination phase, the electron density decreases which leads to the 
decoupling between temperature of the matter and temperature of radiation. 
Chemistry of the early Universe (i.e. $z < 2000$) is the gaseous 
chemistry of the hydrogen, helium, lithium and electrons species. The 
efficiencies of the molecular formation processes is controlled by collisions, 
matter temperature and temperature of the cosmic microwave 
background radiation (CMBR). In the cosmological context  we have metal-free 
gas, and thus the 
formation of $H_2$ is not similar to the formation in the interstellar medium 
by adsorption on the surface of the interstellar grains.
\\
The chemical composition of the primordial gas consists of a mixture of: 
$H$, $H^+$, $H^-$, $D$, $D^+$, $He$, $He^+$, $He^{2+}$, $Li$, $Li^+$, $Li^-$, 
$H_2$, $H_2^+$, $HD$, $HD^+$, $HeH^+$, $LiH$, $LiH^+$, $H_3^+$, $H_2D^+$, $e^-$
, and $\gamma$ which leads to 90 reactions in the chemical network.
\\
It is more convenient to reduce the reactions to only those that are essential
 to accuretaly model the chemistry and to reduce computer times. We 
adopt the concept of {\it minimal model} developped by Abel et al. (1997), and
  focus on the formation of molecular hydrogen. This way we can reduce 
the chemical network to 20 reactions. 
\\
From Saslaw \& Zipoy (1967), and Shchekinov \& Ent\'el (1983) we know that the 
formation of primordial molecular hydrogen is due to the two main reactions:
\begin{equation}
H^- + H \, \longrightarrow \, H_2 + e^-
\end{equation}
\begin{equation}
H_2^+ + H \, \longrightarrow \, H_2 + H^+ .
\end{equation}
These reactions  are coupled with the photo-reactions, the associative, 
recombination and charge exchange reactions (Puy et al. 1993, and Galli 
\& Palla 1998).

\subsection{Equations of evolution}

We consider here the chemical and thermal evolution in the framework of the 
Friedmann cosmological models. The relation between the time $t$ and the 
redshift $z$ is given by:
\begin{equation}
\frac{dz}{dt} \, = \, -H_o (1+z) \, \sqrt{1 + \Omega_o z}
\end{equation}
where $H_o$ is the Hubble constant and $\Omega_o$ the parameter of density of 
the Universe (open Universe $\Omega_o <1$, flat Universe $\Omega_o = 1$ and 
closed Universe for $\Omega_o >1$). The expansion is 
characterized by the adiabatic cooling:
\begin{equation}
\Lambda_{ad} \, = \, 3 n k T_m \, H_o (1 +z ) \, \sqrt{1 + \Omega_o z},
\end{equation}
with the density of matter $n$ and the temperature of matter $T_m$. 
The evolution of the matter density is given by:
\begin{equation}
\frac{dn}{dt} \, = \, -3 n H_o (1 +z ) \, \sqrt{1 + \Omega_o z}.
\end{equation}
and for temperature of radiation:
\begin{equation}
\frac{dT_m}{dt} \, = \, \frac{2}{3nk} \, \Bigl[ 
-\Lambda_{ad} + \Gamma_{compt} + \Psi_{molec} \Bigr],
\end{equation}
where $\Gamma_{compt}$ is the Compton scattering of CMBR photons on 
electrons. Below 4000 K only the rotational levels of $H_2$ can be excited 
(quadrupolar transitions). In the cosmological context Puy et al. (1993) have 
shown that the molecules heat the medium (due to the interactions between 
primordial molecules and the CMBR photons). The thermal molecular function 
$\Psi_{molec}$ (heating minus cooling) is positive in this context.
\\
All these equations are coupled with the set of chemical equations in order 
to calculate the evolution of the abundance of $H_2$. 

\subsection{Evolution of molecular hydrogen}

We consider three sets of parameter for $\Omega_o$ which characterize the 
three particular Universe (open with $\Omega=0.1$, flat with $\Omega_o=1$ and 
closed with $\Omega_o=2$). Moreover we consider two values for the baryonic 
fraction, the lower value obtained with the primordial 
nucleosynthesis $\Omega_\rho =0.02$, and the other which characterizes 
a full baryonic Universe $\Omega_\rho=1$. 
\\
In Fig. 1, we have plotted the different curves for the 
evolution of $H_2$. We see the classical two steps of $H_2$ formation 
(the first step correspond to the $H_2^+$ channel and the second one to the 
$H^-$ channel). After this transient growth $H_2$ abundance becomes constant.

\begin{figure} 
\centerline{\psfig{figure=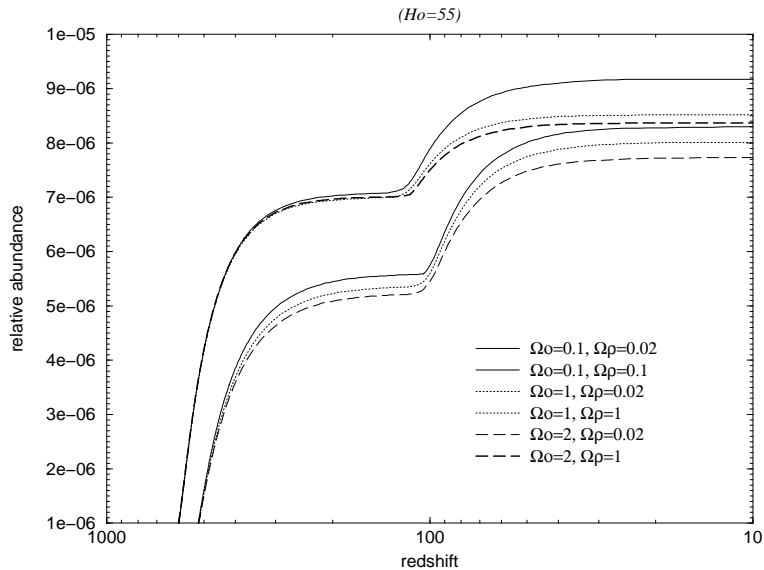,width=10cm,angle=-90}}    
\caption{Evolution of the post-recombination abundance of primordial 
hydrogen for different cosmological models ($\Omega_o$ is the parameter of 
geometry and $\Omega_{\rho}$ the baryonic fraction. The upper curves 
correspond to the lower baryonic fraction ($\Omega_\rho=0.02$), and vice 
versa, the lower curves correspond to the higher baryonic fraction 
($\Omega_\rho = 1$).}
\end{figure}

\section{Cosmological thermal decoupling}

In Fig. 2, we have plotted the ratio between $Tmolec$ which is the temperature 
of matter with primordial molecules and $T_{compt}$ the temperature 
of matter without molecules (we consider in this context only the Compton 
heating). The ratio is closed to unity for the lower value of baryonic 
fraction and close to 2.5 for the higher value. For $ \Omega_b > 0.02$, we
can expected that $T_{molec} \sim 1.25 \times T_{compt}$.

\begin{figure} 
\centerline{\psfig{figure=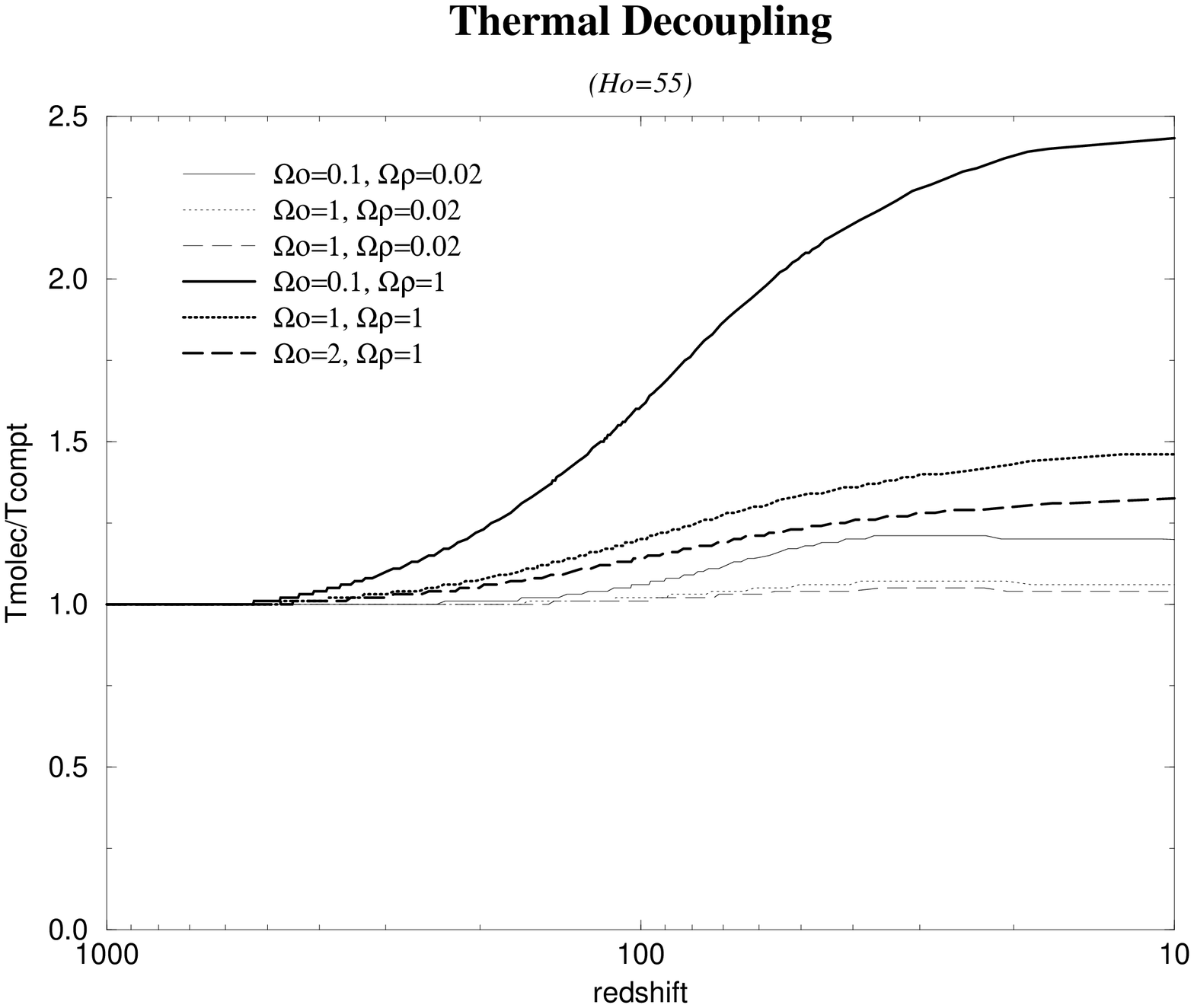,width=10cm,angle=-90}}   
\caption{Evolution of the thermal decoupling for different cosmological 
models ($\Omega_o$ is the parameter of geometry and $\Omega_{\rho}$  
the baryonic fraction).}
\end{figure}

\section{Outlook} 
 
The change of temperature, due to $H_2$, on the thermal decoupling could 
play a role during the transition between the linear regime and the non-linear 
regime (the turn-around point of gravitation collapse), (Puy \& Signore 
1996, Signore \& Puy 1999). The temperature at the turn-around point is 
given by 
$$
T_{turn} \sim \Bigl( \frac{3 \pi}{4} \Bigr)^{4/3} \, T_{compt} 
$$
where $T_{compt}$ is the temperature of matter without the influence of 
molecules (Padmanabhan 1993). Taking into account the influence of the 
molecules, the temperature is 25 per cent more 
important than $T_{compt}$. Thus molecules could give other initial 
conditions for the dynamics of the gravitational collapse than ones predicted 
by the classical theory.
\begin{acknowledgments}
I would like to acknowledge Tom Abel, Lukas Grenacher, Philippe Jetzer and 
Monique Signore for valuable discussions. I thank Francoise Combes for 
organizing such a pleasant conference. This work has been supported by the 
{\it Dr Tomalla Foundation} and by the Swiss National Science Foundation.
\end{acknowledgments}


\begin{thebibliography}{} 

  \bibitem[]{} 
     {\sc Abel T., Anninos P., Zhang Y., Norman M.,} 1997 
     {\it New Astr.} {\bf 2}, 181. 

  \bibitem[]{} 
     {\sc Black J.,} 1988, in: {\it Molecular Astrophysics}, Cambridge 
Univ. Press P. 473.

 \bibitem[]{} 
     {\sc Dalgarno A., Lepp S.,} 1987, in: {\it Astrochemistry IAU 
Symp. 118}, p. 109.

 \bibitem[]{} 
     {\sc Galli D., Palla F.,} 1998 {\it astro-ph/9803315}, 27 Mar 1998.

 \bibitem[]{} 
     {\sc Latter W., Black J.,} 1991 
     {\it ApJ} {\bf 371}, 161.

  \bibitem[]{} 
     {\sc Lepp S., Shull M.,} 1984 
     {\it ApJ.} {\bf 280}, 465. 

 \bibitem[]{} 
     {\sc Olive K.A.,} 1999 
     {\it astro-ph/9901231}, 18 Jan 1999. 

\bibitem[]{}
    {\sc Padmanabhan T.} 1993 in: {\it 
Structure formation in the Universe}, Cambridge University Press.

 \bibitem[]{} 
     {\sc Palla F., Galli D., Silk J.,} 1995
     {\it ApJ} {\bf 451}, 44.             

 \bibitem[]{} 
     {\sc Puy D., Alecian G., Lebourlot J., Leorat J., Pineau des Forets,} 1993
     {\it A\& A } {\bf 267}, 337.

 \bibitem[]{} 
     {\sc Puy D., Signore M.,} 1996
     {\it A\& A } {\bf 305}, 371.

 \bibitem[]{} 
     {\sc Puy D., Signore M.,} 1997
     {\it New Astr.,} {\bf 27}, 622.

 \bibitem[]{} 
     {\sc Puy D., Signore M.,} 1998a
     {\it New Astr.,} {\bf 3 }, 27.

 \bibitem[]{} 
     {\sc Puy D., Signore M.,} 1998b
     {\it New Astr.,} {\bf 3 }, 247.

 \bibitem[]{} 
     {\sc Puy D., Signore M.,} 1999
     {\it New Astr. Rev.,} {\bf 43}, 223.

  \bibitem[]{} 
     {\sc Sarkar S.,} 1996  {\it Rep. Prog. Phys-} {\bf 59}, 1493. 

 \bibitem[]{} 
     {\sc Saslaw W., Zipoy D.,} 1967 
     {\it Nature} {\bf 216}, 967.

 \bibitem[]{} 
     {\sc Shchekinov Y.A., Ent\'el M.B.,} 1983 
     {\it Sov. Astr. Lett.} {\bf 27}, 622.

 \bibitem[]{} 
     {\sc Signore M., Puy D.,} 1999
     {\it New Astr. Rev.,} {\bf 43}, 185.

 \bibitem[]{} 
     {\sc Stancil P., Lepp S., Dalgarno A.,} 1996 
     {\it ApJ} {\bf 458}, 401.

\end{thebibliography}
\end{document}